\documentclass[12pt]{article}
\usepackage{graphicx}
\newcommand{\idk}{\int \frac{{\rm d}^3\bk}{(2\pi)^3}}

\newcommand\nn{\nonumber \\}
\newcommand\beq{\begin{equation}}
\newcommand\eeq{\end{equation}}
\newcommand\beqa{\begin{eqnarray}}
\newcommand\eeqa{\end{eqnarray}}

\newcommand{\ds}[1]{#1 \hspace{-0.5em}/}  
\newcommand\bzeta{\mbox{\boldmath$\zeta$}}

\newcommand\bk{{\bf k}}
\newcommand\bq{{\bf q}}

\def\sla{\slash{\!\!\!} }
\newcommand{\vp}{\mbox{\boldmath $p$}}

\begin{document}

\begin{center}
{\Large\bf Origin of magnetic field in compact stars and magnetic
 properties of quark matter}

\vspace{1cm}

{T. Tatsumi$^1$, T. Maruyama$^2$, K. Nawa$^1$ and E. Nakano$^{3,4}$}

\vspace{0.5cm}

$^1$ Department of Physics, Kyoto University, Kyoto 606-8502, Japan;
 {\it tatsumi@ruby.scphys.kyoto-u.ac.jp}; {\it nawa@ruby.scphys.kyoto-u.ac.jp}\\
$^2$ College of Bioresouce Science, Nihon University,
Fujisawa 252-8510, Japan;{\it tomo@brs.nihon-u.ac.jp}\\
$^3$ Department of Physics, Tokyo Metropolitan University, 
         1-1 Minami-Ohsawa, Hachioji, Tokyo 192-0397, Japan\\
$^4$ Yukawa Institute for Theoretical Physics, Kyoto University, Kyoto
 606-8502, Japan;{\it enakano@yukawa.kyoto-u.ac.jp}
\end{center}

\vspace{1cm}

\begin{abstract}
Microscopic origin of the magnetic field observed in compact stars is
studied in quark matter. Spontaneous spin polarization appears in
high-density region due to the Fock exchange term. On the other hand,
quark matter becomes unstable to form spin density wave in the moderate 
 density region, where restoration of chiral symmetry plays an important 
 role. Coexistence of the magnetism and color superconductivity is
 briefly discussed.

\end{abstract}

\section{Introduction}

Nowadays it is widely accepted that there should be realized various phases of
QCD in temperature ($T$) - density ($\rho_B$) plane. When we emphasize
the low $T$ and high $\rho_B$ region, the subjects are sometimes called
high-density QCD. The main purposes in this field should be 
to figure out the new phases and their properties, and to
extract their symmetry breaking pattern and low-energy excitation modes
there on the basis of QCD. On the other hand, these studies have
phenomenological implications on relativistic heavy-ion collisions and
compact stars like neutron stars or quark stars.

In this lecture we'd like to address magnetic properties of quark matter. 
We shall see various types of magnetic ordering may be expected in quark 
matter at finite density or temperature. They arise due to the quark 
particle-hole ($p-h$) 
correlations in the pseudo-scalar or axial-vector channel.
We first discuss the ferromagnetic phase transition and
then another magnetic feature in the second part.


Phenomenologically the concept of magnetism should be directly related
to the origin of strong magnetic field observed in compact stars \cite{MAG3}; e.g., it
amounts to $O(10^{12}$G) at the surface of radio pulsars. Recently a new
class of pulsars called magnetars has been discovered with super strong
magnetic field, $B_s\sim 10^{14 - 15}$G, estimated from the $P-\dot{P}$
curve \cite{tho04}. First observations  are indirect evidences for 
super strong magnetic field, but 
discoveries of some absorption lines stemming from the cyclotron
frequency of protons have been currently reported \cite{ibr02}. 

The origin of the
strong magnetic field has been a long standing problem since the first
pulsar was discovered \cite{MAG3}. A naive working hypothesis is the
conservation of the magnetic flux and its squeezing during the evolution
from a main-sequence progenitor star to a compact star, $B\propto R^{-2}$ with
$R$ being the radius. Taking the sun as a typical one, we have
$B\sim 10^3$G and $R\sim 10^{10}$cm. If it is squeezed to a typical
radius of usual neutron stars, $R\sim 10$km, the conservation of the
magnetic flux gives $10^{11}$G, which is consistent with the
observations for radio pulsars. However, we find $R\sim 100$m to explain
$B\sim 10^{15}$G observed for magnetars, which leads to a contradiction
since the Schwatzschild radius is $O(1$km) much larger than $R$.   

Since there should be developed hadronic matter inside compact
stars, it would be reasonable to consider a microscopic origin of such
strong magnetic field: ferromagnetism or spin polarization is one of the
candidates to explain it. Makishima also suggested the hadronic origin
of the magnetic field observed in binary X-ray pulsars \cite{mak03}.

When we consider the magnetic-interaction energy by a simple formula,  
$E_{\rm mag}=\mu_iB$ with
the magnetic moment, $\mu_i=e_i/(2m_i)$, we can easily estimate it for
$B=O(10^{15}$G); it amounts to 
several MeV for electrons, while several keV for nucleons and 10 - 100
keV for quarks.
This simple
consideration may imply that strong interaction gives a feasible origin 
for the strong magnetic field, since its typical energy scale is MeV. 
The possibility of
ferromagnetism in nuclear matter has been elaborately studied since the
first 
pulsars were observed, but negative results have been reported so
far \cite{fan01}. In the first part of this lecture, 
we consider its possibility in quark matter from a different point of
view \cite{tat00}.

In the second part we discuss another magnetic aspect in quark matter at
moderate densities, where the QCD interaction is still strong and some
non-perturbative effects still remain. One of the most important
features observed there is restoration of chiral symmetry; the
blocking of $\bar qq$ excitations due to the existence of the Fermi sea
gives rise to restoration of chiral symmetry at a certain density and
many people believe that deconfinement transition also occurs at almost 
the same time. There have been proposed 
various types of the {\it p-h} condensations at moderate densities
\cite{der0, der}, 
in which the {\it p-h} pair in scalar or tensor channel  
has the finite total momentum indicating standing waves 
(the chiral density waves). The instability for the density wave in quark matter was first discussed 
by Deryagin {\it et al.} \cite{der0} at asymptotically high densities 
where the interaction is very weak, 
and they concluded that the density-wave instability prevails over the BCS one 
in the large $N_c$ (the number of colors) limit 
due to the dynamical suppression of colored BCS pairings. 

In general, density waves are favored in 1-D (one spatial dimension) systems 
and have the wave number $Q=2k_F$ according to the Peierls instability 
\cite{peiel1}, 
e.g., charge density waves (CDW) in quasi-1-D metals. 
The essence of its mechanism is the nesting of Fermi surfaces 
and the level repulsion (crossing) of single particle spectra 
due to the interaction for the finite wave number. 
Thus the low dimensionality has a essential role 
to produce the density-wave states. 
In the higher dimensional systems, however, the transitions occur 
provided the interaction of a corresponding ({\it p-h}) channel is strong enough. 
For the 3-D electron gas, 
it was shown by Overhauser \cite{ove} that paramagnetic state is unstable 
with respect to the formation of the static spin density wave (SDW), 
in which spectra of up- and down-spin states are deformed  
to bring about the level crossing 
due to the spin exchange interactions, 
while the wave number does not precisely coincide with $2k_F$ 
because of the incomplete nesting in higher dimension. 

We shall see a kind of spin density wave develops there, in analogy with SDW mentioned above.. 
It occurs along with the chiral condensation 
and is represented by 
a dual standing wave in scalar and pseudo-scalar condensates 
(we have called it `dual chiral-density wave', DCDW). 
DCDW has different features in comparison with 
the previously discussed chiral density waves \cite{der0,der}. One
outstanding feature concerns its magnetic aspect; DCDW induces {\it spin
density wave}.  

\section{Ferromagnetism in QCD}

\subsection{A heuristic argument}

Quark matter bears some resemblance to electron gas interacting with the
Coulomb potential; the
gluon exchange interaction in QCD has some resemblance to the Coulomb 
interaction in QED ,
and color neutrality of quark matter corresponds to charge neutrality of
electron gas under the background of positively charged ions.

It was
Bloch who first suggested a mechanism leading to ferromagnetism of
itinerant electrons \cite{blo}. The mechanism is very simple but largely reflects
the Fermion nature of electrons. Since there works no direct interaction
between electrons as a whole, the Fock exchange interaction gives a
leading contribution.
Then it is immediately conceivable that a most
attractive channel is the parallel spin pair, whereas 
the anti-parallel pair 
gives null contribution (see Eq.~(\ref{nr}) below).
This is nothing but a
consequence of the Pauli exclusion principle: electrons with the same
spin polarization cannot closely approach to each other, which
effectively avoid the Coulomb repulsion. On the other hand a polarized
state should have a larger kinetic energy by rearranging the two Fermi
spheres. Thus there is a trade-off between the kinetic and interaction
energies, which leads to a {\it spontaneous spin polarization (SSP)} 
 or FM at some density.
One of the essential points we learned here is that 
we need no spin-dependent interaction in the original Lagrangian to see
SSP. 

Then it might be natural to ask how about in QCD. We list here some
features of QCD related to this subject. (1) the quark-gluon interaction
in QCD is rather simple, compared with the nuclear force; it is a gauge
interaction like in QED. (2) quark matter should be a color neutral
system and only the $\it Fock~ exchange$ interaction is also relevant like in
the electron system. (3) there is an additional flavor degree of freedom in
quark matter; gluon exchange never change flavor but it comes in through
the generalized Pauli principle. (4) quarks should be treated
relativistically, different from the electron system.

The last feature requires a new definition and formulation of SSP or FM in
relativistic systems since``spin'' is no more a good quantum number in
relativistic theories;
spin couples with momentum and its direction changes during the motion.  
It is well known that the Pauli-Lubanski vector $W^\mu$ is the four vector to 
represent the
spin degree of freedom in a covariant form; the spinor of the free Dirac 
equation is the eigenstate of the operator,
\beq
W\cdot a=-\frac{1}{2}\gamma_5\ds{a}\ds{k},
\label{aab}
\eeq
where a 4-axial-vector $a^\mu$ s.t.
\beq
{\bf a}=\bzeta+\frac{\bk(\bzeta\cdot\bk)}{m(E_k+m)}, 
~a^0=\frac{\bk\cdot\bzeta}{m}~~~
\label{ac}
\eeq
with the axial vector $\bzeta$. We can see that $a^\mu$ is reduced to a
three 
vector $(0, \bzeta)$
in the rest frame, where we can allocate $\bzeta=(0,0,\pm 1)$ to spin ``up'' and ``down'' states.
Thus we can still use $\bzeta$ to specify the two intrinsic polarized states
even in the general Lorentz frame.

The Fock exchange interaction, $f_{{\bf k}\zeta,{\bf q}\zeta'}$, between
two quarks is then given by 
\beq
 f_{{\bf k}\zeta,{\bf q}\zeta'}
=g^2\frac{m}{E_k}\frac{m}{E_q}
\frac{2}{9m^2}[2m^2-k\cdot q-m^2 a\cdot b]\frac{1}{(k-q)^2},
\label{ce}
\eeq
in the lowest order, where the spin dependent term renders 
\beqa
a\cdot b&=&-\frac{1}{m_q^2}\left[
-(\bk\cdot\bzeta)(\bq\cdot\bzeta')
+m^2\bzeta\cdot\bzeta'\right.\nonumber\\
&+&\left\{m(E_k+m)(\bzeta\cdot\bq)(\bzeta'\cdot\bq)
+m(E_q+m)(\bzeta'\cdot\bk)(\bzeta\cdot\bk)\right.\nonumber\\
&+&\left.\left.(\bk\cdot\bq)(\bzeta\cdot\bk)
(\bzeta'\cdot\bq)\right\}/(E_k+m)(E_q+m)\right].
\label{cg}
\eeqa
It exhibits a complicated spin-dependent structure arising from the
Dirac four spinor, while it is reduced to a simple form,
\beqa
-\frac{2}{9}g^2\frac{1+\bzeta\cdot \bzeta'}{({\bf
k}-{\bf q})^2}
\label{nr}
\eeqa
in the non-relativistic limit as in the electron system. Eq.~(\ref{nr})
clearly shows why parallel spin pairs are favored, while we cannot see
it clearly in the relativistic expression (\ref{cg}). We have
explicitly demonstrated that the ferromagnetic phase should be realized at 
relatively low density region \cite{tat00}: quark matter is spontaneously
polarized at low densities like the electron gas. The phase transition
is of first order and the critical density is around the nuclear density
for quark mass $m_q\simeq 300$MeV and coupling constant $\alpha_c\sim
2.2$, which are used in the MIT bag model. It should be interesting to
see a recent reference \cite{nie}, where the author also found the
ferromagnetic phase transition at low densities within the perturbative
QCD calculation beyond the lowest-order diagram.

\subsection{Self-consistent calculation}

If we understand FM or magnetic properties of quark matter more deeply,
we must proceeds to a self-consistent approach, like the Hartree-Fock
theory,  
beyond the previous perturbative argument.

We begin with an OGE action:
\beqa
    I_{int}=-g^2\frac{1}{2}\int{\rm d^4}x \int{\rm d^4}y
 \left[\bar{\psi}(x)\gamma^\mu \frac{\lambda_a}{2} \psi(x)\right]
D_{\mu \nu}(x,y)
 \left[\bar{\psi}(y)\gamma^\nu \frac{\lambda_a}{2} \psi(y)\right], 
\label{ogeaction}
\eeqa
where $D^{\mu\nu}$ denotes the gluon propagator. 
By way of the mean-field approximation, we have 
\beq
 I_{MF}=\int \frac{{\rm d}^4 p}{(2 \pi)^4} 
                \bar\psi(p) G_A^{-1}(p) \psi(p). 
\label{mfield}
\eeq
The inverse quark Green function $G^{-1}(p)$ involves various self-energy 
(mean-field) terms, of which we only keep the color singlet
particle-hole mean-field 
$V(p)$,
\beq
G_A(p)^{-1}= \sla{p}-m+\sla{\mu}+V(p).
\label{gainv}
\eeq
Taking into account the lowest diagram, we can then write down the 
self-consistent equations for the mean-field, $V$:
\beqa
-V(k)=(-ig)^2 \int \frac{{\rm d}^4p}{i(2\pi)^4} \{-iD^{\mu \nu}(k-p)\} 
      \gamma_\mu \frac{\lambda_\alpha}{2} \{-iG_A(p)\} 
      \gamma_\nu \frac{\lambda_\alpha}{2}  
\label{self1}.
 \eeqa
\begin{figure}[h]
\begin{center}
\includegraphics[width=6cm,clip]{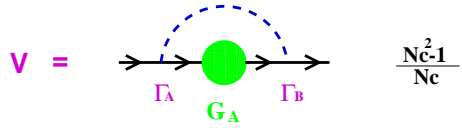}
\end{center}
\caption{Graphical interpretations of the equation (\ref{self1}) 
with coefficient in front of R.H.S. given by $N_c$.}
\end{figure}

Applying the Fierz transformation for the OGE action (\ref{ogeaction})
we can see that 
there appear the color-singlet scalar, pseudo-scalar, vector and axial-vector
self-energies by the Fock exchange interaction.   
In general we must take into account these self-energies in $V$,
$V=U_s+\gamma_5U_{ps}+\gamma_\mu U_v^\mu+\gamma_\mu\gamma_5U_{av}^\mu$
with the mean-fields $U_i$. Here we retain only $U_s, U_v^0, U_{av}^3$
in $V$ and suppose that others to be vanished.
 
\medskip
\centerline {
\hfill\vbox{\hsize15.5cm\hrule\hbox to 15.5cm{\vrule\hfill\vbox to 1.5cm{\hsize15cm
\medskip\noindent
{\bf Problem}: 
Show that there is no tensor mean-field as a consequence
of chiral symmetry in (\ref{ogeaction}).
\vfill}\hfill\vrule}\hrule}\hfill}

According to the above assumptions and considerations 
the mean-field $V$ in Eq.(\ref{gainv}) renders 
\begin{equation}
V = \gamma_3 \gamma_5 U_A, ~~~U_A\equiv U_{av}^3 , 
\end{equation}
with the axial-vector mean-field $U_A$
\footnote{Since the scalar and vector mean-fields only renormalize the mass
and the quark-number chemical potential, respectively, we discard
them here for simplicity.}
. 

The poles of $G_A(p)$, $\det$$G^{-1}_A$($p_0$$=$$\epsilon_n$)$=$$0$,
give the single-particle energy spectrum:
\beqa
&& \epsilon_n=\pm \epsilon_\pm \\
  && \epsilon_{\pm}= \sqrt{{\bf p}^2+{\bf U}_A^2+m^2 \pm 2 
                        \sqrt{m^2 {\bf U}_A^2+({\bf p}\cdot {\bf U}_A)^2 }},
\label{eig}
\eeqa
where the subscript $\pm$ in the energy spectrum represents spin degrees of
freedom, and the dissolution of the degeneracy
corresponds to the {\it exchange splitting} of 
different ``spin'' states \cite{yoshi}.

There are two Fermi seas with different volumes for a given quark number 
due to the exchange splitting in the energy spectrum. 
The appearance of the rotation symmetry breaking term, $\propto {\bf
p}\cdot {\bf U}_A$ in the energy
spectrum implies deformation of the Fermi sea: thus 
rotation symmetry is violated in the momentum space as well as the
coordinate space, $O(3)\rightarrow O(2)$. Accordingly the Fermi sea
of majority quarks exhibits a ``prolate'' shape ($ F^-$), while that 
of minority quarks an ``oblate'' shape ($F^+$) as seen Fig.~2.
\begin{figure}[h]
\begin{center}
\includegraphics[width=10cm,clip]{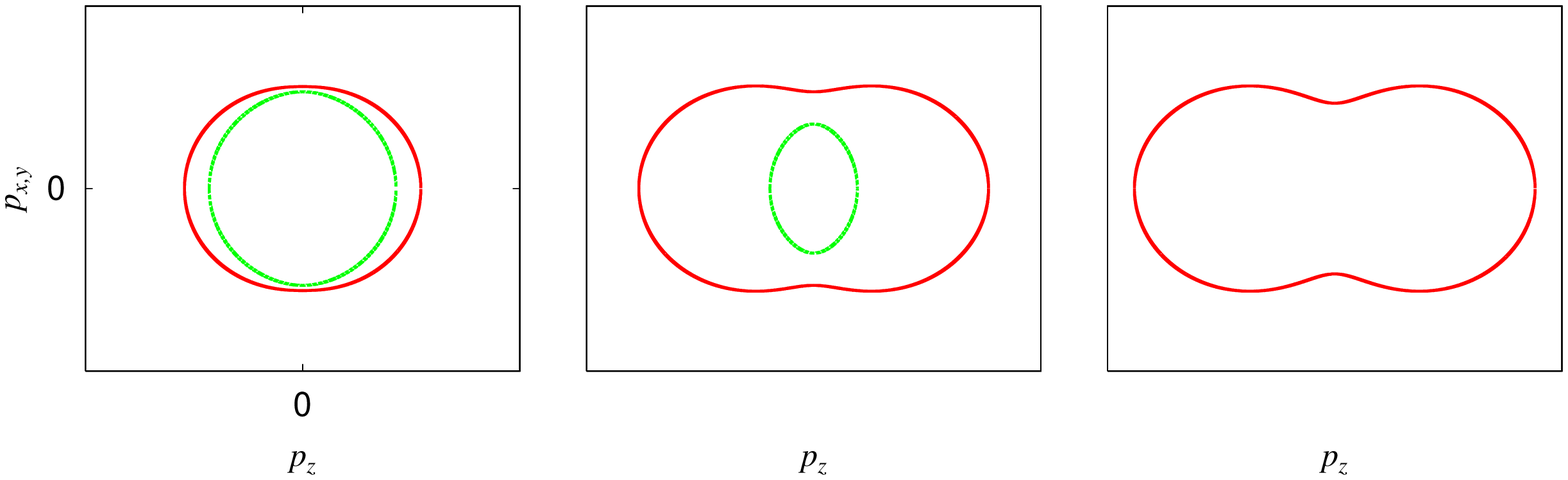}
\end{center}
\caption{Modification of the Fermi sea as $U_A$ is increased from left
 to right. The larger Fermi sea ($F^-$) takes a prolate shape, while
 the smaller one ($F^+$) an oblate shape for a given $U_A$. In the
 large $U_A$ limit (completely polarized case), $F^+$ disappears as 
in the right panel.}
\end{figure}

Here we demonstrate some numerical results; we replaced the original OGE
by the ``contact'' interaction,
$
D^{\mu\nu}\rightarrow -g^{\mu\nu}/\Lambda^2.
$
Then the self-consistent equation can be written as
\beq
  U_A=-\frac{N_c^2-1}{4N_c}\tilde{g}^2 \int \frac{{\rm d}^3 p}{(2\pi)^3} 
\sum_s\theta(\mu-\epsilon_s(\vp))
\frac{U_A +s \beta_p}{\epsilon_s(\vp)}, 
\label{UA1}
\eeq
within the ``contact'' interaction, $\tilde g^2\equiv g^2/\Lambda^2$.
Note that the expression for $U_A$,
   Eq.~(\ref{UA1}), is nothing but the simple sum of the expectation
   value of the spin operator over the Fermi seas.

\subsection{Phase diagram in temperature-density plane} 
We will present the phase diagram in the three-flavor case under mainly
two conditions \cite{nak05}:
the chemical equilibrium condition (CEC) $\mu_u=\mu_d=\mu_s$ and 
the charge neutral condition without electrons (CNC) $\rho_u=\rho_d=\rho_s$, 
where quark masses are taken as $m_u=m_d=5$MeV and $m_s=150-350$MeV, 
i.e., $\mu_{s}=\sqrt{\mu_{u,d}^2+m_s^2-m_{u,d}^2}$ for $T=0$.  
In both conditions, 
since the spin polarization caused by the axial-vector mean-field is fully enhanced by the quark masses,   
choice of the current quark mass affects the results,  
especially, the strange quark mass is large, and therefore has essential effect on the spin polarization. 
To get the phase diagram, 
we use the thermodynamic potential $\Omega$ within the mean-field approximation, 
\begin{eqnarray}
\Omega&=&
-N_c \sum_{u=\pm1}~\sum_{s=\pm}~ \sum_{i=u,d,s} \idk 
 T\log
\left\{ \exp\left[ -\frac{\epsilon_s(\bk,m_i,U_A)-u\mu_i}{T}\right]+1 \right\}
\nn
&&-N_c \sum_{s=\pm}~ \sum_{i=u,d,s} \idk \epsilon_s(\bk,m_i,U_A) +\frac{U_A^2}{4 \tilde{g}^2}.
\end{eqnarray}
The second term is the vacuum contribution and can be regularized by
,e.g.,  the proper-time method. 
We can see that the potential $U_A$ reproduces the self-consistent
equation 
Eq.~(\ref{UA1}) in the three-flavor case. 

\begin{figure}[h]
\begin{center}
\includegraphics[height=4.5cm]{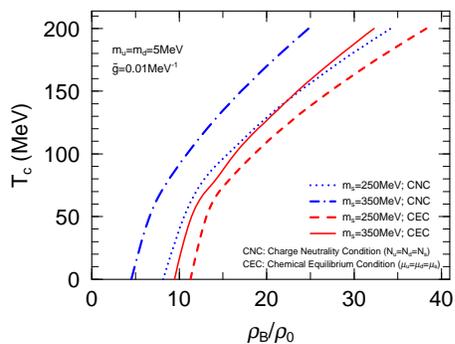}
\end{center}
\caption{Phase diagram for two cases of strange quark masses, $250, 350$MeV.}
\label{pd3}
\end{figure}
Fig.~\ref{pd3} shows the resulting phase diagram under the two conditions mentioned above; 
CEC and CNC.  

The figure shows the critical temperature (the Curie temperature) for a given baryon density, 
CNC tends to facilitate the system being the spin polarization than CEC.  
This is because CNC holds the larger strange-quark density in comparison with CEC. 

\begin{figure}
\begin{center}
\includegraphics[height=4.5cm]{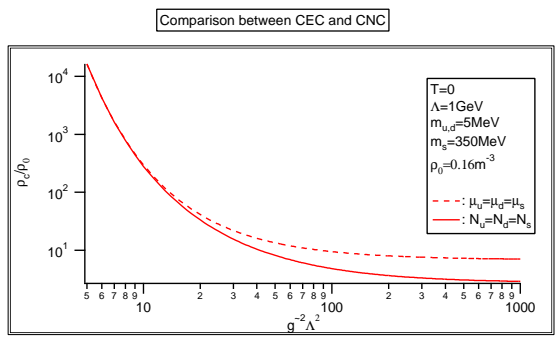}
\end{center}
\caption{Critical density as a function of $\tilde{g}$ under two conditions: CEC and CNC.}
\label{comp1}
\end{figure}

Since the axial-vector mean-field comes from the Fock exchange
interaction  
and causes a kind of particle-hole condensation,  
which is enhanced by the Fermi sea contribution,   
there exists a critical density for a given coupling constant $\tilde{g}$.   
We show the critical density by varying the effective coupling constant $\tilde{g}$ in Fig.~\ref{comp1}.  
The critical density lowers with the larger coupling strength, 
and this tendency is enhanced in the case of CNC. 
The result also indicates that even for the weak-coupling regime in QCD,  
the spin polarization may appear in sufficiently large densities at sufficiently low temperature.    

\subsection{Color magnetic superconductivity}

If FM is realized in quark matter, it might be in the color
superconducting (CSC) phase \cite{bai,ber}.
In this section we briefly discuss a possibility of the coexistence of FM and
CSC, which we call {\it Color magnetic superconductivity} \cite{nak}. 
In passing, it
would be worth mentioning the corresponding situation in condensed
matter physics. Magnetism and superconductivity (SC) have been two major
concepts in condensed matter physics and their interplay has been
repeatedly discussed \cite{MagSup1}. Very recently some materials have been observed to
exhibit the coexistence phase of FM and SC \cite{MagSup2}.
In our case we shall
see somewhat different features, but the similar aspects as well.

We begin with the OGE action (7) and extend it to include the quark
pairings. We consider the quark pairings on the same Fermi surfaces
$F^\pm$ by using the eigenspinors $\phi_s, s=\pm 1$ with the energy 
(\ref{eig});
the pairing function then reads
\begin{eqnarray}
\Delta({\vp})=\sum_{s=\pm} \tilde{\Delta}_s({\vp}) B_s({\vp}),~~~
B_s({\vp})=\gamma_0 \phi_{-s}({\vp}) \phi_{s}^\dagger({\vp}).
\label{delta}
\end{eqnarray}   
The structure of the gap
function (\ref{delta}) is inferred from a physical consideration of a
 quark pair as in the usual BCS theory: 
we consider here the quark pair on each Fermi surface 
with opposite momenta, ${\bf p}$ and $-{\bf p}$ so that they result in a linear combination of
$J^\pi=0^-, 1^-$.  $\tilde\Delta_s$ is still a matrix in the color-flavor space. Taking into account the property that the most attractive channel of 
the OGE interaction is  
the color antisymmetric ${\bar 3}$ state, the quark pair must be in the flavor singlet 
state. Thus we can  choose the form of the gap function as 
\beq
\left(\tilde\Delta_s\right)_{\alpha\beta;ij}=\epsilon^{\alpha\beta 3}\epsilon^{ij}\Delta_s
\label{gapfn}
\eeq
for the two-flavor case (2SC), where $\alpha,\beta$ denote the color indices and
$i,j$ the flavor indices. In refs. \cite{nak} we can see that the gap
functions $\Delta_s$ have a polar angle ($\theta$) dependence on the
Fermi surfaces: both the gap functions have nodes at poles
($\theta=0,\pi$) and take the maximal values at the vicinity of equator
($\theta=\pi/2$), keeping the relation, $\Delta_- \geq \Delta_+
$. This feature is very similar to $^3 P$ pairing in liquid $^3$He or
nuclear matter \cite{leg,NM3P}; actually we can see our pairing function
 Eq.~(\ref{gapfn}) to exhibit an 
effective $P$ wave nature by a genuine relativistic effect by the Dirac spinors. 
As a consequence, e can say that FM and CSC barely interfere with each
other \cite{nak}.

\section{Dual chiral density wave}
\subsection{Chiral symmetry restoration and 
Instability of the directional mode}

We consider here another type of density wave described as a dual
standing wave in the scalar and pseudo-scalar mean-fields \cite{tat04}.
It is well known that chiral symmetry is spontaneously broken (SSB) 
due to the quark-anti-quark condensate in the vacuum and at low densities; 
both the scalar and pseudo-scalar densities always 
reside on the chiral circle of the finite modulus, and any chiral 
transformation by a chiral angle $\theta$ shifts each value on the circle, 
while the QCD Lagrangian is invariant for constant $\theta$.

The spatially variant chiral angle $\theta({\bf r})$ 
represents the degree of freedom 
of the Nambu-Goldstone mode in the SSB 
vacuum. The dual chiral density wave (DCDW) 
is described by such a chiral angle 
$\theta({\bf r})$.
When the chiral angle has some space-time 
dependence, there should appear extra terms in the effective potential:
one trivial term
is the one describing the quark and DCDW   
coupling due to 
the non-commutability of $\theta({\bf r})$ with the kinetic
(differential) operator in the Dirac operator.  
Another one is nontrivial and comes from the 
quantum effect: the energy spectrum of the quark is modified in the
presence of $\theta({\bf r})$ and thereby the vacuum energy has an 
additional term, $\propto (\nabla\theta)^2$ in the lowest order. 
This can be regarded as an appearance of the kinetic term for
DCDW through the vacuum polarization \cite{sug}.  

\subsection{DCDW in the NJL model}

Taking the Nambu-Jona-Lasinio (NJL) model as a simple but nontrivial 
example, we explicitly demonstrate that quark matter 
becomes unstable for a formation of DCDW above 
a critical 
density; the NJL model has been  
recently used as 
an effective model of QCD, embodying spontaneous breaking of chiral symmetry 
in terms of quark degree of freedom \cite{kle}
\footnote{We can see that the OGE interaction gives the same 
form after the Fierz transformation in the zero-range limit \cite{nak}}
.
We shall explicitly see the DCDW state exhibits a ferromagnetic property.

We start with the NJL Lagrangian with $N_f=2$ flavors and $N_c=3$ colors,
\beq
{\cal L}_{NJL}
=\bar\psi(i\ds{\partial}-m_c)\psi+G[(\bar\psi\psi)^2+
(\bar\psi i\gamma_5\mbox{\boldmath$\tau$}\psi)^2],
\label{njl}
\eeq
where $m_c$ is the current mass, $m_c\simeq 5$MeV.
Under the Hartree approximation, we linearize Eq.~(\ref{njl}) by partially 
replacing the bilinear quark fields by their expectation values 
with respect to the ground state. 

In the 
usual treatment to  
study the restoration of chiral symmetry at finite density,  
authors implicitly discarded the speudo-scalar mean-field, while  
this is justified only for the vacuum of a definite parity.
We assume here the following mean-fields,
\beqa
\langle\bar\psi\psi\rangle&=&\Delta\cos(\bf q\cdot\bf r) \nonumber\\
\langle\bar\psi i\gamma_5\tau_3\psi\rangle&=&\Delta\sin(\bf q\cdot\bf r), 
\label{chiral}
\eeqa
and others vanish
\footnote{It would be interesting to recall that the DCDW  
configuration is similar to pion condensation in high-density nuclear 
matter within the $\sigma$ model, 
considered by Dautry and Nyman (DN)\cite{dau,kut}, where $\sigma$ and $\pi^0$ 
meson condensates take the same form as Eq.~(\ref{chiral}). 
The same configuration has been also assumed for non-uniform chiral phase 
in hadron matter by the use of the Nambu-Jona-Lasinio model \cite{sad}.
However, DCDW is by no means the pion condensation but should
be directly considered as particle-hole and particle-antiparticle quark
condensation 
in the deconfinement phase. }
. Accordingly, we define a new quark field $\psi_W$ by the Weinberg 
transformation,
\beq
\psi_W=\exp[i\gamma_5\tau_3 {\bf q\cdot r}/2 ]\psi,
\label{wein}
\eeq
to separate the degrees of freedom of the amplitude and phase of  
DCDW in the Lagrangian. In terms of the new field the effective Lagrangian 
renders
\beq
{\cal L}_{MF}=\bar\psi_W[i\ds{\partial}-M-1/2\gamma_5\tau_3\ds{q}]\psi_W
-G\Delta^2,
\label{effl}
\eeq
where we put $M\equiv -2G\Delta$ and $q^\mu=(0, {\bf q})$, taking the
chiral limit ($m_c=0$). 
The form given in (\ref{effl}) appears to be the same as 
the usual one,  except the axial-vector field 
generated by the wave vector of DCDW; the {\it amplitude} of DCDW 
produces the dynamical quark mass in this case.  
We shall see 
the wave vector ${\bf q}$ is related to the magnetization: the 
{\it phase} of DCDW induces the magnetization.

Using Eq.~(\ref{effl}), 
we can find a 
spatially uniform solution,
$\psi_W=u_W(p)\exp(i{\bf p\cdot r})$, 
with the eigenvalues,
\beq
E_p^{\pm}=\sqrt{E_{p}^{2}+|{\bf q}|^2/4\pm \sqrt{({\bf
p}\cdot{\bf q})^2+M^{2}|{\bf q}|^2}},~~~E_p=(M^2+|{\bf p}|^2)^{1/2}
\label{energy}
\eeq
for positive-energy (valence) quarks with different spin polarizations 
(c.f. (\ref{eig})). 

\subsection{Thermodynamic potential}

The thermodynamic potential is given as
\beqa
\Omega_{\rm total}&=&\gamma\int\frac{d^3p}{(2\pi)^3}
\left[(E^-_p-\mu)\theta_-+(E^+_p-\mu)\theta_+\right]
-\gamma\int\frac{d^3p}{(2\pi)^3}\left[E^-_p+E^+_p\right]
+M^2/4G\nonumber\\
&\equiv&\Omega_{\rm val}+\Omega_{\rm vac}+M^2/4G .
\label{therm}
\eeqa
where $\theta_\pm=\theta(\mu-E^\pm_p)$, $\mu$ is the chemical potential and 
$\gamma$ the degeneracy factor $\gamma=N_fN_c$. The first term 
$\Omega_{\rm val}$ is the 
contribution by the valence quarks filled up to the chemical potential, 
while the second term $\Omega_{\rm vac}$ is the vacuum 
contribution that is apparently divergent. We shall see both contributions
are {\it indispensable} in our discussion.  
Once $\Omega_{\rm total}$ is properly evaluated, the equations to be solved to 
determine the optimal values of $\Delta$ and $q$ are 
\beq
\frac{\delta\Omega_{\rm total}}{\delta\Delta}
=\frac{\delta\Omega_{\rm total}}{\delta q}=0.
\label{self}
\eeq

Since NJL model is not renormalizable, we need some regularization procedure 
to get a meaningful finite value for the vacuum contribution
$\Omega_{\rm vac}$, which can be recast in the form,
\beq
\Omega_{\rm vac}=i\gamma\int\frac{d^4p}{(2\pi)^4}{\rm trln}S_W,
\eeq
with use of the propagator $S_W=(\ds{p}-M-1/2\tau_3\gamma_5\ds{q})^{-1}$.
Since the energy spectrum
is no more rotation symmetric, we cannot apply the usual momentum cut-off 
regularization (MCOR)
scheme to regularize $\Omega_{\rm{vac}}$
\footnote{We would like to stress that the regularization should be, at
least, independent of $\Delta$ and $q$. Otherwise it is inconsistent
with Eq.~(\ref{self}). }
. Instead, we adopt the 
proper-time 
regularization (PTR) scheme \cite{sch}. 
Introducing the proper-time variable $\tau$, we eventually find
\beq
\Omega_{\rm vac}=\frac{\gamma}{8\pi^{3/2}}\int_0^\infty
\frac{d\tau}{\tau^{5/2}}
\int^\infty_{-\infty}\frac{dp_z}{2\pi}\left[
e^{-(\sqrt{p_z^2+M^2}+q/2)^2\tau}
+e^{-(\sqrt{p_z^2+M^2}-q/2)^2\tau}\right]-\Omega_{\rm ref},
\label{j}
\eeq
which is reduced to the standard formula \cite{kle} 
in the limit $q\rightarrow 0$.

The integral with respect to the proper time $\tau$ is not well defined as 
it is,     
since it is still divergent 
due to the $\tau\sim 0$ contribution.
Regularization proceeds by replacing the lower bound of the integration range 
by $1/\Lambda^2$, which corresponds to the momentum cut-off in the 
MCOR scheme.

Now we examine a possible instability of quark matter with respect to formation of DCDW. In the following we first consider the sign change of the 
curvature of $\Omega_{\rm total}$ at the origin ({\it stiffness} parameter), 
$\beta$.
Expanding $\Omega_{\rm vac}$ with respect to $q$ up to $O(q^2)$, we find
\beqa
\Omega_{\rm vac}
&=&\Omega_{\rm vac}^0+\frac{\gamma\Lambda^2}{16\pi^2}J(M^2/\Lambda^2)q^2+O(q^4)
\nonumber\\
&\equiv &\Omega_{\rm vac}^0+\Omega^{mag}_{\rm vac}+O(q^4),
\label{l}
\eeqa
where $J(x)$ is a universal function,
$
J(x)=-x{\rm Ei}(-x),
$
with the exponential integral ${\rm Ei}(-x)$.
$\Omega^{mag}_{\rm vac}$ is the pure magnetic contribution and provides a 
kinetic term ($\propto (\nabla\theta)^2$) for DCDW
. It originates from 
a vacuum 
polarization effect in the 
presence of DCDW and gives a 'repulsive' (positive) contribution, 
so that the vacuum is stable against 
formation of DCDW. Note that it gives a null contribution 
in case of $M=0$ 
, irrespective of $q$, as it should be. 

The coefficient of $q^2$ term in 
$\Omega^{mag}_{\rm vac}$, the vacuum stiffness parameter, $\beta_{vac}$, 
has a definite 
physical meaning. Since the pion decay 
constant $f_\pi$ is mass dependent and is given in terms of $J(x)$ 
within the NJL model \cite{kle},
$\beta_{vac}$ can be written as
$\beta_{vac}=\frac{1}{2}f_\pi^2$.

For given $\mu, M$ and $q$ we can evaluate the valence contribution 
$\Omega_{\rm val}$ using 
Eq.~(\ref{energy}), 
but its general formula  is very complicated. However, it may be sufficient to 
consider the small $q$ case for our present purpose. 
Then the thermodynamic potential can be expressed as
$
\Omega_{\rm val}=\epsilon_{\rm val}(q)-\mu\rho_{\rm val}(q),
$
where $\epsilon_{\rm val}(q)$ and $\rho_{\rm val}(q)$ are the energy density and the quark-number 
density, respectively.  
Expanding $\epsilon_{\rm val}(q)$ up to the $O(q^2)$ 
we find
\beq
\epsilon_{\rm val}(q)
\simeq\epsilon_{\rm val}^0+\frac{\gamma}{8\pi^2}M^2q^2
\left[\frac{\mu}{\sqrt{\mu^2-M^2}}-{\rm ln}\frac{\mu+\sqrt{\mu^2-M^2}}{M}
\right]+O(q^4)
\label{x}
\eeq
with 
$
\epsilon_{\rm val}^0=\gamma/8\pi^2\left[\mu\sqrt{\mu^2-M^2}
\left(2\mu^2-M^2\right)
-M^4{\rm ln}(\mu+\sqrt{\mu^2-M^2}/M)\right]$ for normal quark matter
without DCDW.
Thus the valence contribution can be finally written as
\beqa
\Omega_{\rm val}&=&\Omega_{\rm val}^0
-\frac{\gamma}{8\pi^2}M^2q^2H(\mu/M)+O(q^4)\nonumber\\
&\equiv& \Omega_{\rm val}^0+\Omega_{\rm val}^{mag}+O(q^4)
\label{xb}
\eeqa 
up to $O(q^2)$, where $H(x)={\rm ln}(x+\sqrt{x^2-1})$ and $\Omega_{\rm val}^0
=\epsilon_{\rm val}^0-\mu\rho_{\rm val}^0$ with 
$\rho^0_{\rm val}=\frac{\gamma}{3\pi^2}(\mu^2-M^2)^{3/2}$ for normal
quark matter.
The valence stiffness parameter then reads
\beq
\beta_{val}=-\frac{\gamma}{8\pi^2}M^2H(\mu/M)
\eeq
Since the function $H(x)$ is always positive and accordingly 
$\beta_{val}\leq 0$,
the magnetic term $\Omega_{\rm val}^{mag}$ always gives a negative energy and 
approaches to zero as $M\rightarrow 0$ (triviality).  

\begin{figure}[h]
\begin{center}
\includegraphics[width=0.3\textwidth,keepaspectratio]{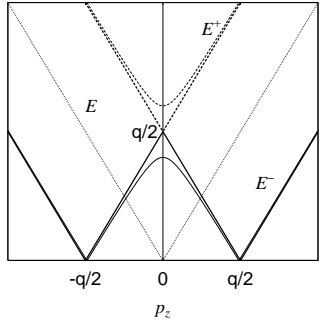}
\end{center}
 \caption{Energy spectra for ${\bf p}_{\perp}=0$. $E^\pm$ with $M=0$ 
(thick solid and dashed lines).
$\tilde E^\pm$ with the 
definite chirality is also shown for comparison (dotted line). 
We can see there is a degeneracy of $E^\pm$ 
at $p_z=0$ for $M=0$, while it is resolved by the mass 
(thin solid and dashed lines).}
\label{level}
\end{figure}

We 
may easily understand why the valence quarks always favor the formation of 
DCDW. 
First, consider the energy spectra for massless quarks (see Fig.~\ref{level}). 

As is 
already discussed, our theory becomes trivial in this case and we find two 
spectra   
\beq
E^\pm_p=\sqrt{p_\perp^2+(|p_z|\pm q/2)^2}, ~~~{\bf p}_\perp=(p_x,p_y, 0),
\label{zeom}
\eeq
which are essentially equivalent to $E^\pm_p=|{\bf p}|$. 

\medskip
\centerline {
\hfill\vbox{\hsize15.5cm\hrule\hbox to 15.5cm{\vrule\hfill\vbox to 1.5cm{\hsize15cm
\medskip\noindent
{\bf Problem}: 
Construct the eigenstates of chirality $\gamma_5$ by taking the linear
combination of $\psi_W^\pm$. Their spectra are given by $\tilde E^\pm=\sqrt{p_\perp^2+(p_z\pm q/2)^2}$.
\vfill}\hfill\vrule}\hrule}\hfill}

There 
is a level crossing at ${\bf p}={\bf 0}$. Once the mass term is taken into 
account this degeneracy is resolved and the exchange splitting appears there. 
Hence it causes an energy gain, if $q=O(2\mu)$; we can 
see that this  
mechanism is very similar to that of SDW by Overhauser \cite{der,ove}.

Using Eqs.~(\ref{therm}), (\ref{l}), (\ref{xb}) we write the thermodynamic 
potential as 
\beq
\Omega_{\rm total}=\Omega_{NJL}+\beta q^2+O(q^4)
\label{xf}
\eeq
with the total stiffness parameter $\beta=\beta_{vac}+\beta_{val}$ and the usual 
NJL expression without DCDW,
$
\Omega_{NJL}=\Omega_{\rm vac}^0(M)+\Omega_{\rm val}^0(M)+M^2/4G.
$
The dynamical quark mass $M$ is 
given by the equation, $\partial\Omega_{\rm total}/\partial M=0$;
At the order of $q^0$ the dynamical quark mass $M^0$ is determined by the equation,
$
\left.\partial\Omega_{NJL}/\partial M\right|_{M^0}=0.
$
Since $M-M^0=O(q^2)$, DCDW onsets 
at a certain density where the total stiffness parameter $\beta$  
becomes negative:  the critical chemical 
potential $\mu^{cr}$ is determined by the equation,
\beq
\beta
=\frac{1}{2}f_\pi^2-\frac{\gamma}{8\pi^2}\left(M^0\right)^2H(\mu^{cr}/M^0)=0.
\label{y}
\eeq
In the limit $M^0\rightarrow 0$, we find 
$
\tilde\mu^{cr}=1/2e^{-\gamma_E/2}\simeq 0.375...
$
with $\gamma_E$ being Euler's constant and $\tilde\mu=\mu/\Lambda$ in the PTR 
scheme.
 Note that this is only a {\it sufficient} condition for 
the 
existence of DCDW phase, and we can {\it never} exclude the 
possibility of the first order phase transition or metamagnetism \cite{tat00,blo}.    
Actually, we shall see that DCDW occurs as a first-order phase transition.

\subsection{First-order phase transition}

\begin{figure}
\vspace*{-0cm}
\begin{center}
\includegraphics[height=9cm, angle=-90]{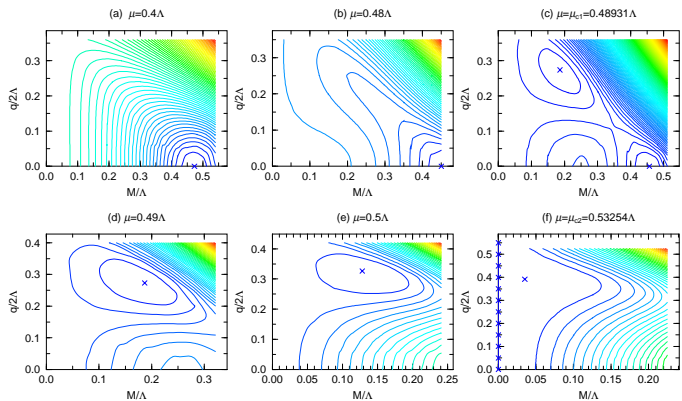}
\end{center}
\caption{Contours of $\Omega_{\rm total}$ at $T=0$ are shown in $M-q$ plane 
as the chemical potential increases, (a) $\rightarrow$ (f).  
The cross in each figure denotes the absolute minimum.}
\label{cp1}
\end{figure}

The magnitudes of $M$ and $q$ are obtained from the minimum 
of the thermodynamic potential ~(\ref{therm}) for $T=0$.  
Fig.~\ref{cp1} shows the contours of $\Omega_{\rm tot}$ in $M$-$q$ plane  
as the chemical potential increases,   
where the parameters are set as $G\Lambda^2=6$ and $\Lambda=850$ MeV, 
which are not far from those for the vacuum ($\mu=0$)  \cite{kle}.  

The crossed points denote the absolute minima. 
There are two critical chemical potential $\mu=\mu_{c1}, \mu_{c2}$:    
for the lower densities (Fig.~\ref{cp1}(a)-(b)) 
the absolute minimum resides at the point $(M\neq 0, q=0)$ 
indicating the SSB phase. 
At $\mu=\mu_{c1}$ (Fig.~\ref{cp1}(c)) 
the potential has the two absolute minima at $(M\neq 0, q=0)$ and $(M\neq 0, q\neq 0)$, 
showing the first-order transition to the DCDW phase   
which is stable for $\mu_{c1}<\mu<\mu_{c2}$ (Fig.~(\ref{cp1})d-e). 
At $\mu=\mu_{c2}$ (Fig.~\ref{cp1}(f))   
the axis of $M=0$ and a point $(M\neq 0, q\neq 0)$ become minima, 
the system undergoes the first-order transition again to the chiral-symmetric phase.    

\begin{figure}[h]
\begin{minipage}{0.48\textwidth}
\begin{center}
\includegraphics[height=4.5cm]{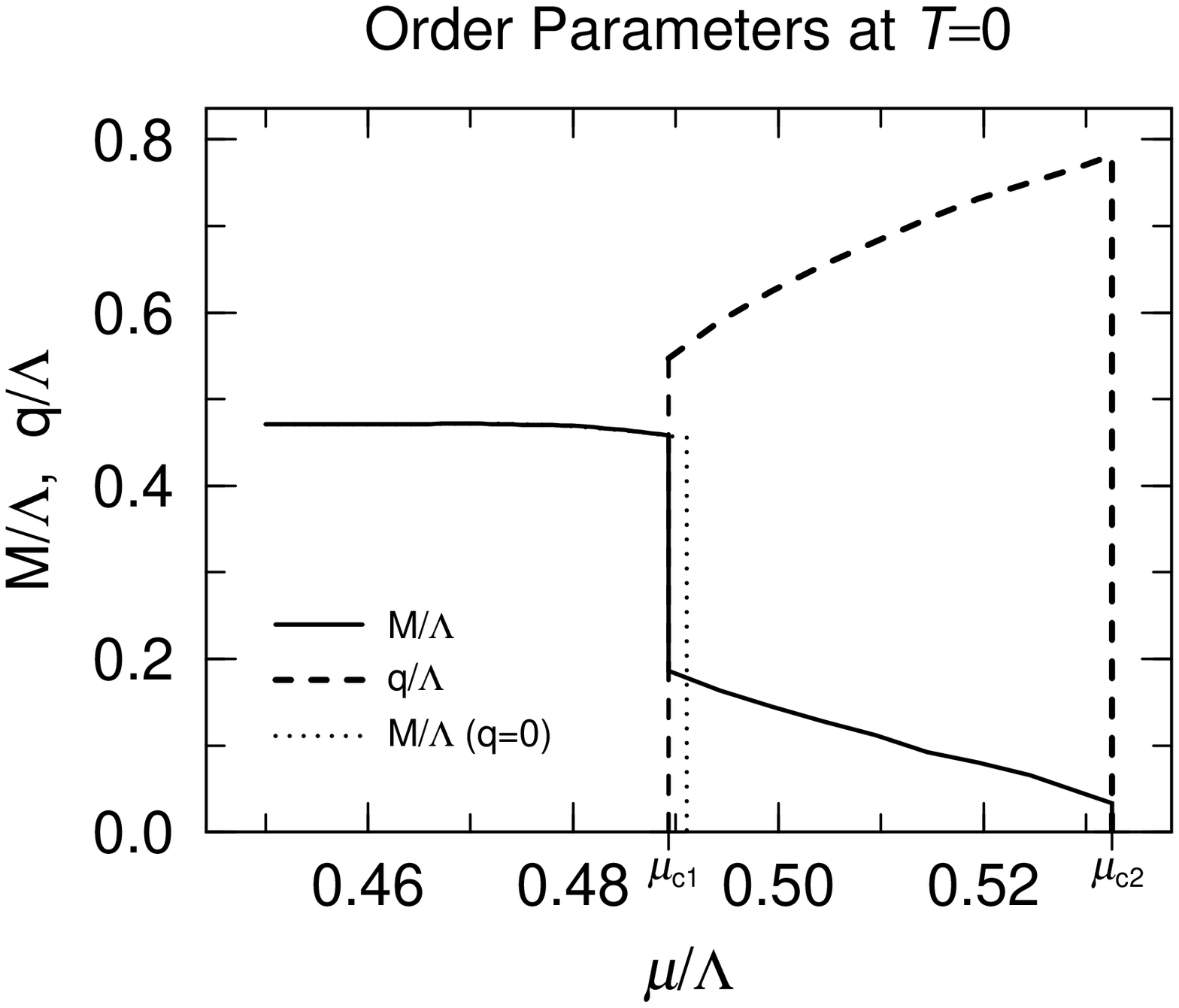}
\end{center}
\caption{Wave number $q$ and the dynamical mass $M$ are plotted 
as functions of the chemical potential at $T=0$. 
Solid (dotted) line for $M$ with (without) the density wave, 
and dashed line for $q$.}
\label{op1}
\end{minipage}
\hspace{\fill}
\begin{minipage}{0.48\textwidth}
\begin{center}
\includegraphics[height=4.5cm]{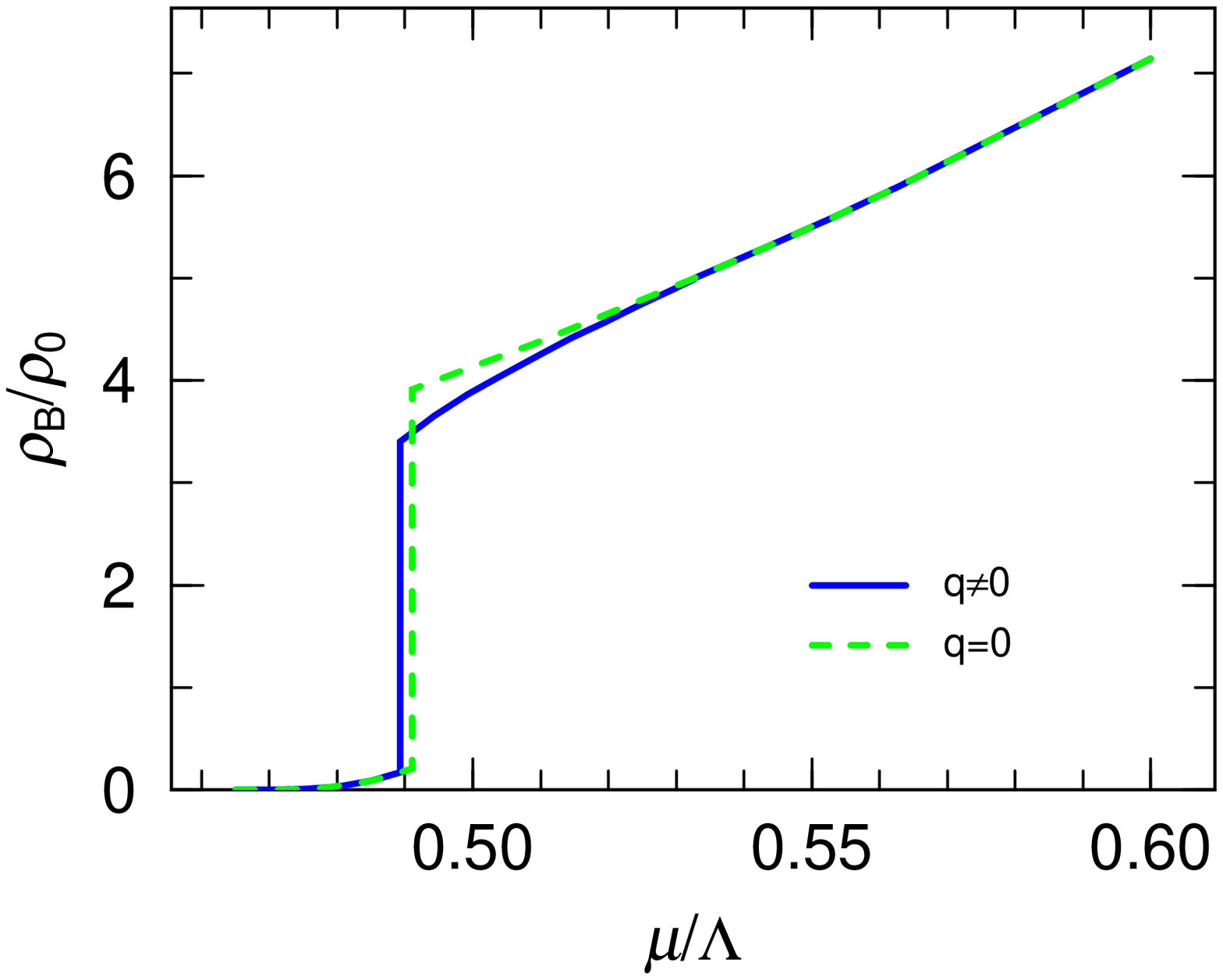}
\end{center}
\vspace*{-0.5cm}
\caption{Baryon number density as a function of $\mu$. 
$\rho_0=0.16 {\rm fm}^{-3}$: the normal nuclear density.}
\label{rhoCDW}
\end{minipage}
\end{figure}

Fig.~\ref{op1} shows the behaviors of the order-parameters $M$ and $q$ 
as functions of $\mu$ at $T=0$, 
where that of $M$ without DCDW is also shown for comparison. 
It is found from the figure that 
DCDW develops at finite range of $\mu$ ($\mu_{c1}\le\mu\le\mu_{c2}$), where  
the wave number $q$ increases with $\mu$, which value 
is smaller than twice of the Fermi momentum 
$2k_F$($\simeq 2\mu$ for free quarks) 
due to the higher dimensional effect; 
the nesting of Fermi surfaces is incomplete in the present 3-D system. 
Actually, the ratio of the wave number and the Fermi momentum 
(at normal phase $q=M=0$) becomes $q/k_F=1.17-1.47$ 
for the baryon-number densities
$\rho_b/\rho_0=3.62-5.30$ where the DCDW is stable (see Fig.~8).  

\subsection{Magnetic properties}

The mean-value of the spin operator is given by 
\beq
\bar s_z=\frac{1}{2}u^\dagger_W\Sigma_z u_W
=\frac{1}{2}\frac{q/2\pm\beta}{E^\pm_p}+{\rm vac},
\label{spin}
\eeq
with $\beta=\sqrt{p_z^2+m^2}$, where "vac" means the vacuum contribution.
First note that the integral of $\bar s_z$ over the Fermi seas should be proportional to $q$, and 
the solution with $q\neq 0$ seems to imply FM. However, we can show that PTR 
gives the 
vacuum (the Dirac sea) contribution oppositely to cancel the total mean-value 
of the spin operator, which is consistent with Eq.~(\ref{self}). 
Instead we can see that the magnetization spatially 
oscillates, 
\beq
M_z\equiv \langle \bar q\sigma_{12}q\rangle=\langle\gamma_0\sigma_{12}\rangle \cos({\bf q}\cdot{\bf r}),
\eeq
with 
\beq
\langle\gamma_0\sigma_{12}\rangle=\int_{F^+-F^-}\frac{d^3p}{(2\pi)^3}\frac{2M}
{\sqrt{M^2+p_z^2}},
\eeq
which means a kind of spin density wave \cite{gru}.

\subsection{Phase diagram at $T-\mu$ plane}
To complete the phase diagram  
we derive the thermodynamic potential at finite temperature 
in the Matsubara formalism. 
The partition function for the mean-field Hamiltonian is given by  
\begin{eqnarray}
Z_\beta &=&
\int {\it D}\bar{\psi} {\it D}\psi 
\exp \int_0^\beta d\tau \int d^3r ~
 \left\{ \bar{\psi}\left[
i \tilde{\partial} +M \exp\left(i\gamma_5 {\bf q} \cdot {\bf r}\right) 
-\gamma_0 \mu \right]\psi 
-\frac{M^2}{4 G} \right\}\nonumber\\ 
&=& 
\prod_{\bk,n,s=\pm} 
\left\{ (i \omega_n+\mu)^2-E_s^2(\bk) \right\}^{N_fN_c}
\times \exp\left\{-\left( \frac{M^2}{4 G}\right)V\beta \right\},
\end{eqnarray}
where $\beta=1/T$, 
$\tilde{\partial}\equiv -\gamma_0 \partial_\tau+i {\bf \gamma}\nabla$ and 
$\omega_n$ the Matsubara frequency. 
Thus the thermodynamic potential $\Omega_\beta$ is obtained,   
\begin{eqnarray}
\Omega_\beta (q,M)\!\!\!
&=&\!\!\!-T \log{Z_\beta(q,M)}/V \nn
\!\!\!&=&\!\!\!-N_fN_c\!\!\idk\sum_s\left\{
T\log{\left[ e^{-\beta\left(E_s(\bk)-\mu \right)}\!\!+\!\!1 \right] 
      \left[ e^{-\beta\left(E_s(\bk)+\mu \right)}\!\!+\!\!1 \right]}
\!\!+\!\!E_s(\bk) \right\}\!\!+\!\!\frac{M^2}{4 G}\label{tpot}. 
\end{eqnarray}

\begin{figure}
\begin{center}
\includegraphics[height=6.3cm]{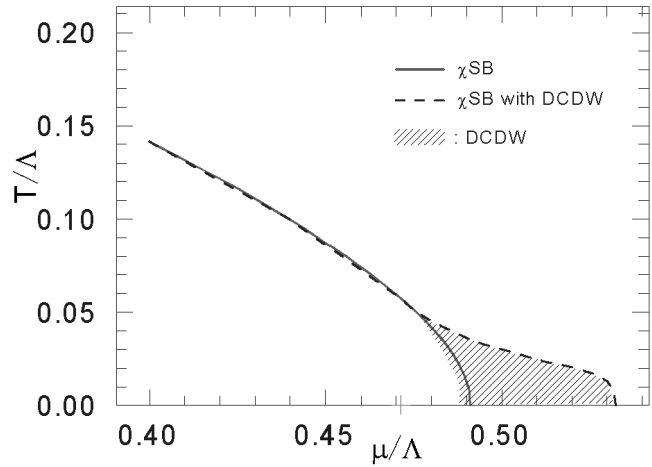}
\end{center}
\caption{Phase diagram obtained 
from the thermodynamic potential ~(\ref{tpot}). 
The solid (dashed) line shows the chiral-restoration line 
in the presence (absence) of DCDW. 
The shaded area shows the DCDW phase. }
\label{pd1}
\end{figure}

From the absolute minima of the thermodynamic potential (\ref{tpot}), 
it is found that 
the order parameters at $T\neq0$ behave similarly to those at $T=0$ 
as a function of $\mu$,  
while the chemical-potential range of the DCDW at finite temperature, 
$\mu_{c1}(T)\le \mu \le\mu_{c2}(T)$,  
gets smaller as $T$ increases. 
We show the resultant phase diagram in Fig.~(\ref{pd1}), 
where the ordinary chiral-transition line is also given. 
Comparing phase diagrams with and without $q$,  
we find that the DCDW phase emerges in the area 
(shaded area in Fig.~(\ref{pd1}))
which lies just outside the boundary of the ordinary chiral transition.
We thus conclude that the DCDW is induced by finite-density contributions, 
and has an effect to expand the chiral-condensed phase ($M\neq 0$) 
toward low temperature and high density region. 

\section{Summary and Concluding remarks}

We have seen some magnetic aspects of quark matter: ferromagnetism at 
high densities and spin density wave at moderate densities within the 
zero-range approximation for the interaction vertex. These look to follow 
the similar development about itinerant electrons: Bloch mechanism at 
low densities and spin density wave at high densities by Overhauser.

With the OGE interaction, we have seen ferromagnetism in quark matter at
low densities (Sect.2.1)
. It would be worth mentioning that another 
study with higher-order diagrams also supports it. These studies suggest an opposite 
tendency to the one using the zero-range interaction (Sects
2.2,2.3). Note that we can also see the 
same situation for the itinerant electrons; the Hartree-Fock calculation based 
on the infinite-range Coulomb interaction favors ferromagnetism at low 
density region, while the Stoner model, which introduces the zero-range effective 
interaction instead of the Coulomb interaction, gives ferromagnetism at high 
densities. So we must carefully examine the possibility of ferromagnetim in 
quark matter by taking into account the finite-range effect.   
 
We have seen that dual chiral desnity wave (DCDW) appears at a certain density 
and develops at moderate densities (Sect. 3). It occurs as a result of the interplay  
between the $\bar qq$ and particle-hole correlations. The phase 
transition is of weakly first order, and the restoration of chiral symmetry is 
delayed compared with the usual scenario. Note that there is no soft
pions in this phase, and phason may appear istead as a Nambu-Goldstone boson.

For the discussion of DCDW, we have seen the remarkable 
roles of the Fermi sea and the Dirac sea: the former always favors  
DCDW, while the latter works against it. The similar 
situation also appears about the magnetic property of quark matter. The mean 
value of the spin operator over the Fermi seas of valance quarks always gives a
finite value 
in the presence of DCDW, which is a kind of 
ferromagnetism, but the vacuum contribution given by 
the Dirac seas completely cancels it. As a result there is no net spin 
polarization in this case, but we have seen magnetization spatially oscillates 
instead (spin density wave). This is one of the typical examples in which the nonrelativistic 
picture is qualitatively different from the relativistic one by the vacuum 
effect.

It would be ambitious to give a scenario based on ferromagnetism of quark 
matter, which can explain the hierarchy of the magnetic field observed 
in three classes of neutron stars, magnetars, radio pulsars and recycled 
millisecond pulsars. It would be also interesting to study some implication 
of spin density wave or phason on the magnetic properties of compact stars.

\section*{Acknowledgments}

This work is partially supported by the Grant-in-Aid for the 21st Century COE
``Center for the Diversity and Universality in Physics '' from 
the Ministry of Education, Culture, Sports, Science and
Technology of Japan. It is also partially supported by the Japanese 
Grant-in-Aid for Scientific
Research Fund of the Ministry of Education, Culture, Sports, Science and
Technology (13640282, 16540246).

\end{document}